**Structural Changes and Ferroelectric Properties of BiFeO$_3$-PbTiO$_3$ Thin Films Grown via a Chemical Multilayer Deposition Method**

Shashaank Gupta[1], Ashish Garg[2], Dinesh Chandra Agrawal[1], Shuvrajyoti Bhattacharjee[3] and Dhananjai Pandey[3]

[1]Materials Science Programme

[2]Materials and Metallurgical Engineering

Indian Institute of Technology, Kanpur 208016, India

[3]School of Materials Science and Technology

Institute of Technology, Banaras Hindu University, Varanasi 221005, India

**Abstract:**

Thin films of (1-x)BiFeO$_3$-xPbTiO$_3$ (BF-xPT) with x ≈ 0.60 were fabricated on Pt/Si substrates by chemical solution deposition of precursor BF and PT layers alternately in three different multilayer configurations. These multilayer deposited precursor films upon annealing at 700°C in nitrogen show pure perovskite phase formation. In contrast to the equilibrium tetragonal structure for the overall molar composition of BF:PT::40:60, we find monoclinic structured BF-xPT phase of $M_A$ type. Piezo-force microscopy confirmed ferroelectric switching in the films and revealed different normal and lateral domain distributions in the samples. Room temperature electrical measurements show good quality ferroelectric hysteresis loops with remanent polarization, $P_r$, of up to 18 μC/cm$^2$ and leakage currents as low as 10$^{-7}$ A/cm$^2$.

## 1. Introduction:

BiFeO$_3$ (BF), a multiferroic material showing co-existence of ferroelectric and magnetic ordering in the same phase, is the most studied multiferroic material since the report of existence of very large polarization in epitaxial BiFeO$_3$ thin films.[1] It shows a ferroelectric-paraelectric transition at ~1103 K[2] and G-type antiferromagnetic-paramagnetic transition at ~643 K with an incommensurate spiral magnetic ordering.[3] It possesses a rhombohedrally distorted perovskite structure with space group R3c at room temperature.[4] Recently, it has been shown to exhibit a series of other transitions: existence of a β-phase between 820 and 925°C followed by the presence of a γ-phase between 925-933°C[5]; a spin-glass transition at 120 K and a ferromagnetic transition at 5 K.[6] Studies on single crystals have shown very high values of polarization, close to 100 μC/cm$^2$.[7] However, polycrystalline BiFeO$_3$ ceramics and films have often been plagued by high electrical conductivity, due to which hysteresis loops appear leaky. This is believed to be due to the presence of various secondary phases and valence fluctuation at the Fe-site (Fe$^{3+}$ and Fe$^{2+}$) during processing.[8,9]

Solid-solution of BF with insulating ABO$_3$ perovskite oxides, such as PbTiO$_3$ (PT)[10] and BaTiO$_3$[11], has been suggested as one of the alternatives to improve the electrical resistance of the material. Among these, the mixed (1-x)BiFeO$_3$–xPbTiO$_3$ (BF-xPT) system is very interesting as it not only forms a continuous solid solution but also exhibits a morphotropic phase boundary (MPB)[12] with a narrow composition width Δx ≈ 0.03 in which the tetragonal (T) and rhombohedral ( R) phases coexist.[13] Outside the MPB region, only the T and R phases are observed for x≥0.31 and x≤0.27, respectively.[13] This solid solution system is also attractive because of the high ferroelectric Curie temperature in the range 763 to 1100K[12], and unusually large tetragonality [η = (c/a) - 1 = 0.187 *i.e.* 18.7%] for x = 0.31[13,14] which is three times that in PbTiO$_3$. These features have led to many studies on BF-xPT films in the past few years where films have shown good low temperature ferroelectric properties.[15,16,17,18,19,20] However, none of these studies report good quality room temperature ferroelectric hysteresis loops, especially at frequencies below 1 kHz, presumably due to high intrinsic leakage.

Here, we report on the structural changes and insulation resistance of BF-xPT thin films with x ≈ 0.60 obtained by a multilayer deposition approach involving spin coated precursor layers of BiFeO$_3$ and PbTiO$_3$ of varying thickness and number of bi-layers on Pt/Si substrates. In contrast to the expected tetragonal structure as reported in the bulk samples of the same composition[12], the structure of these films is found to be monoclinic. These films show reasonably high remanent polarization at room temperature with excellent leakage characteristics.

## 2. Experimental Details:

A 0.15M BF precursor solution was prepared by first dissolving high purity bismuth nitrate pentahydrate and iron (III) nitrate nonahydrate in glacial acetic acid and 2-methoxyethanol, respectively, followed by mixing of the two solutions and further addition of acetic anhydride and a few drops of ethanolamine for stabilization. This solution was subsequently stirred for 3 hours. Similarly, a 0.15M PT precursor solution was prepared by dissolving lead (II) acetate trihydrate in glacial acetic acid followed by drop-wise addition of titanium (IV) butoxide (stabilized with acetyl acetonate) and subsequent stirring for 3 hours. The precursor PT and BF layers were alternately deposited on platinized-Si substrates by spin coating at a speed of 4000 rpm for 30 seconds. After each coating, the films were dried for 12 minutes at 360°C followed by deposition of next layer. Finally after completion of deposition, whole stack was annealed at 700°C for 1 h in nitrogen atmosphere. Three different types of layered configurations were synthesized: BFPT16, BFPT8 and BFPT4: BFPT16 contains sixteen bi-layers of BF and PT of

~10 and ~20 nm each, BFPT8 contains eight bi-layers of ~20 nm thick BF and ~40 nm thick PT, and BFPT4 contains four bi-layers of ~40 nm thick BF and ~80 nm thick PT. The overall stack thickness was maintained at ~480-500 nm for all the three configurations. In all configurations, PT layer was deposited before BF layers. The average composition of these films corresponds to x=0.60.

Thickness of the films was measured by Tencor Alpha X-100 Profilometer. Thermoelectron ARL X'tra X-Ray Diffractometer was used for collecting powder X-ray diffraction (XRD) data. The package Fullprof (J. Rodriguez-Carvajal, FULLPROF, Laboratory Leon Brillouin _CEA-CNRS_CEA/Saclay, 91191 Gif sur Yvette Cedex, France, 2006) was used for Le-Bail refinements using the XRD data in the 2θ range of 15 to 120 degrees. In the refinements, the data in the 2θ range of 36 to 43.3 and 85 to 88 were excluded due to strong texture effects and the substrate peaks. Further, coexistence of Pt (FCC) substrate peaks was also taken into account in the refinements.

Scanning electron microscope (SEM, Zeiss SUPRA™ 40 VP) was used for microstructure and compositional studies. Contact mode piezoresponse force microscopy (PFM) studied were carried out using Scanning Probe Microscope (NT-MDT Solver). Magnetic measurements were carried out using a Vibrating sample magnetometer (ADE-DMS EV-7VSM) at room temperature. 200 μm diameter platinum electrodes were sputter deposited on the films to facilitate the electrical measurements. The ferroelectric properties of the films were measured by Radiant Precision LC ferroelectric tester, respectively. Current-voltage characteristics were measured by Keithley 6517A Electrometer.

## 3. Results and Discussion:

*3.1 Structural characterization*

Grazing angle X-Ray diffraction (XRD) patterns of the three samples in the 2θ range of 15 to 80 degrees are shown in Figure 1 (a). The patterns do not show the presence of individual peaks of PT and BF suggesting intermixing of these during post-deposition heat treatment and formation of a solid-solution. This also indicates that the XRD peaks in Fig. 1 (a) belong to BF-0.60PT solid-solution phase only. In order to verify the compositional variations across the thickness of the films, cross-sectional SEM was performed on all the samples and a representative micrograph of a BFPT16 sample (having thickest bi-layers) is shown in Figure 2. The micrograph does not show any contrast between individual BF and PT layers suggesting intermixing between them. The image also shows that films thickness was reasonably uniform. Further, EDAX measurements showed that the films contain approximately 60% PT. Although these observations appear to be in agreement with the XRD results showing a single phase material, local compositional gradients cannot be ruled out due to insufficient resolution of SEM image and errors associated with EDAX measurements.

For the average nominal composition BF-0.60PT of these films, the structure should be tetragonal in the P4mm space group. However, the patterns shown in Fig. 1 (a) do not match with tetragonal BF-xPT, but show some resemblance with the rhombohedral structure of BF-xPT. For example, the 110 pseudocubic peak, which occurs around 2θ = 32.5° is supposed to be a doublet for the tetragonal structure with the stronger 101 peak occurring on the lower 2θ angle side while the weaker 110 peak on the higher side. The observed 110 pseudocubic peak profile does not show such a splitting. On the contrary, it shows an asymmetric tail towards the lower 2θ side suggesting the presence of a weak reflection overlapping with the stronger one on the higher 2θ angle side. This particular feature supports rhombohedral structure but then the 100 pseudocubic peak near 2θ = 23° should have been a singlet. The fact that this peak shows a pronounced asymmetric tail on the lower 2θ side suggests the presence of two overlapping peaks which in turn rules out rhombohedral structure as well. Monoclinic phases in the Cm and Cc space groups have been recently reported in PZT ceramics.[21,22] It has also been shown that the so-called rhombohedral compositions in R3m and R3c space group of PZT are also monoclinic in Cm and Cc space groups, respectively.[22,23]

To correctly predict the structure of the BF-0.60PT films, we further carried out the refinement of the XRD data by LeBail technique considering rhombohedral and monoclinic structures in the R3m and Cm space groups respectively. We did not consider R3c and Cc space groups, as these are not distinguishable from R3m and Cm space groups, respectively, using X-ray powder diffraction data from thin films, because of the extremely low intensities of the characteristic superlattice reflections resulting from the antiphase rotation (tilting) of oxygen octahedra about one of the pseudocubic <111> directions in the neighboring perovskite cells. In fact, these superlattice reflections are not discernible in Fig. 1 (a). The fits for a few selected profiles, as obtained from the full pattern LeBail refinement in the 2θ range 15 to 120 degrees, are shown in Fig. 1 (b) for the two structural models. The upper and lower rows of the vertical bars below the fitted profiles correspond to the perovskite and Pt peaks. It is evident from this figure that the rhombohedral space group can not account for the observed profiles, as it gives rise to huge mismatch between the observed and calculated profiles. The fit for the monoclinic structure, on the other hand, is very good, as indicated by significantly lower $\chi^2$ values also. The refined cell parameters of the monoclinic phase for the films (BFPT4, 16, 18) are given in Table 1. These cell parameters suggest that this monoclinic phase is of $M_A$ type (in the notation of Vanderbilt and Cohen)[24] similar to that reported in PZT.[21,22] A monoclinic structure of $M_A$ type has previously been reported for epitaxial BF films[1] and its presence in thin films has been attributed to the strain effects.[1]

*3.2 Electrical Characterization*

The contact topography image obtained on BFPT16 sample shows an average grain size of about 200 nm (see Fig 3 (a)). Similar microstructures and grain sizes were observed in other samples. All the samples showed piezo-response, as confirmed by recording images without and with voltage applied to the AFM tip. PFM contrast appeared only after voltage was applied to the tip (see Fig. 3 (b)). PFM measurements on the three samples revealed the presence of both normal and lateral domains with different distributions over the sample surface analyzed, as shown in Fig. 3 ((c) and (d)) for BFPT16. The measurements were made at 15 V bias and 150 kHz modulation frequency. The presence of both normal and lateral domain structures is indicative of ferroelectric switching in the films. The brighter regions in these images are representative of high piezo-activity whilst darker regions indicate regions low piezo-activity. In the normal piezoresponse images (figure 3 (c)), the bright and dark contrast regions represent the downward and upward deflections caused by domain polarization along the z-axis, while in the lateral piezoresponse images shown in Fig. 3 (d), the bright and dark areas correspond to the lateral deflections of the domains.[25] The medium contrast regions represent grains possessing weak piezoelectric response due to the deviation of the polarization vector away from the direction normal to the film plane.

Room temperature electrical measurements were conducted on all the three samples. Figure 4 shows the ferroelectric hysteresis loops for the three samples. The measurements were made at a frequency of 1 kHz under a bipolar electric field. The loops were not leaky as evidenced by their saturated nature. The samples were able to withstand fields above 800 kV/cm, suggesting highly insulating behaviour of all the films. The polarization (2P$_r$) values were ~36 μC/cm$^2$ for BFPT16 and decreased to ~ 30 μC/cm$^2$ for BFPT4 and ~16.5 μC/cm$^2$ for BFPT8. Saturated hysteresis loops have been previously seen in undoped BF-xPT system but at temperatures much lower than room temperature at which the hopping conductivity is drastically suppressed.[15,19] Good quality room temperature ferroelectric hysteresis loops with higher remnant polarization have been reported in bulk La-doped BF-xPT,[26] but not in pure, undoped BF-xPT films. The presence of good quality loops in our samples can be attributed to our processing technique (multilayer deposition) which also led to a structural change observed in the present study. This M$_A$ type monoclinic structure of BFPT would have polarization vector which can rotate continuously on a (110) type pseudocubic plane in contrast to the tetragonal structure whose polarization vector is constrained along <100> pseudocubic direction. The ability of the polarization vector to rotate continuously on a given plane offers easier alignment of dipoles upon switching.[21,22,24] The different polarizations of three samples can be attributed to minor changes in the structural parameters of three samples (see Table 1), possibly related to subtle compositional changes across the film thickness.

The results of *dc* leakage measurements on these three samples are shown in Figure 5. The figure shows that the values of leakage current are below $10^{-5}$ A/cm$^2$ in our multilayer BFPT16 samples and below $10^{-6}$ A/cm$^2$ for both BFPT4 and BFPT8 samples at 300 kV/cm. These leakage current values are significantly lower as compared to the reported room temperature values for most polycrystalline BiFeO$_3$ thin films[27,28] and most BF-xPT films reported thus far.[15,17,19,29,30] Figure 5 shows that at high fields, a relationship of the form $\log_{10}J \propto E^{1/2}$ exists suggesting bulk dominated conduction *i.e.* Poole-Frenkel effect. This highly insulating behavior appears to stem from the incorporation of PT into BF via a layered approach which reduces the large scale transport of the carriers or charged defects. We find the significant role of PT precursor layer, which was always deposited first on the Pt/Si substrate. The advantageous role of such a PT-buffered layer has already been demonstrated by us[20] even for the normal chemical solution deposited (without multilayer deposition approach) BF-xPT films where we showed that when BFPT film is deposited without first growing a PT layer, the properties remain inferior. PT, intrinsically being a better insulator than BF, may reduce the transport of charge carriers to the electrodes and hence improve the leakage characteristics. High resistivity of the samples also suggests that samples do not possess significant amount of conducting secondary phases or mixed valence states of Fe.

*3.3 Magnetic Measurements*

Preliminary studies also reveal a saturated magnetic hysteresis loop depicting high magnetization in these films. For instance, Figure 6 shows the room temperature magnetic hysteresis loop of a BFPT 16 sample showing remnant magnetization ($M_r$) of ~0.13 $\mu_B$/f.u. which is considerably larger than the reported values for single crystal $BiFeO_3$.[7] Previous studies have shown that finite magnetization is achievable in BF by melting the spiral spin cycloid via application of large enough magnetic fields[31], substrate induced constraints in epitaxial $BiFeO_3$ thin films,[1] or by means of chemical substitutions.[12,32] In the present samples, high magnetic moment could be a possibility by either of thin film clamping effect, presence of PT in the solid solution or structural change or their combination. Initial investigations on BF-xPT suggest that the magnetization initially increases up to~ 50% PT followed by a decrease at higher BF contents. This indirectly suggests that magnetization is likely to be due to intrinsic effects e.g. ordering of $Fe^{3+}$ and $Fe^{+2}$ ions in the lattice. However, more work is needed to understand whether or not magnetic impurities like γ-Fe$_2$O$_3$, not detectable in the XRD data, are responsible for the observed magnetization.

## 4. Conclusions

In summary, BF-0.60PT films with high insulation resistance and high polarization at room temperature were fabricated via a multilayer approach using chemical solution deposition method. The structure of the resulting thin films was found to be monoclinic in contrast to the tetragonal structure predicted by the overall stoichiometry for bulk samples. Piezoforce microscopy confirmed the presence of ferroelectric switching in the samples. Maximum remnant polarization is achieved in case of bi-layer thickness of ~ 30 nm and is possibly due to improved intermixing of the layers. The samples also show switching up to very high fields (~800 kV/cm) and leakage currents as low as $10^{-7}$ A/cm$^2$. Our work demonstrates that it is possible to make highly insulating BiFeO$_3$-xPbTiO$_3$ films via a multilayer chemical solution deposition approach.

Acknowledgements

AG thanks Department of Science and Technology and Defense Research and Development Organization (both Government of India) for Ramanna fellowship and the financial support.

2  Table 1.    Refined cell parameters for the three films (BFPT16, BFPT8 & BFPT4) obtained

3  by LeBail refinement with Cm space group.

| Sample | $a_m$ (Å) | $b_m$ (Å) | $c_m$ (Å) | β (degree) |
|---|---|---|---|---|
| BFPT16 | 5.456(1) | 5.525(9) | 4.0373(3) | 90.72(2) |
| BFPT8 | 5.409(1) | 5.5056(7) | 4.0385(3) | 90.94(2) |
| BFPT4 | 5.385(1) | 5.488(2) | 4.037(1) | 91.18(2) |

**List of Figures**

Figure 1.  (a) Powder x-ray diffraction (XRD) patterns of BFPT16, BFPT8 and BFPT4. The inset shows zoomed 100 and 110 pseudocubic profiles for BFPT16 (b) The LeBail fits for the XRD data of BFPT16 using R3m and Cm space groups. The dots represent the observed data while the continuous line represents the calculated profiles. The bottom plot is the difference profile. The upper and lower set of bars above the difference profiles correspond to Bragg peaks for BFPT and Pt respectively.

Figure 2.  Cross-sectional back-scattered SEM image of a BFPT16 sample

Figure 3.  (a) Contact mode AFM image of a BFPT16 film (b) PFM image showing the difference in contrast upon application of voltage (c) Normal domain distribution and (d) Lateral domain distribution in BFPT 16 sample.

Figure 4.  Ferroelectric hysteresis loops of BFPT4, BFPT8 and BFPT16 samples.

Figure 5.  Leakage characteristics of the BFPT4, BFPT8 and BFPT16 samples.

Figure 6.  Room temperature magnetic hysteresis measurements on BFPT16 sample (M: magnetization and H: applied field)

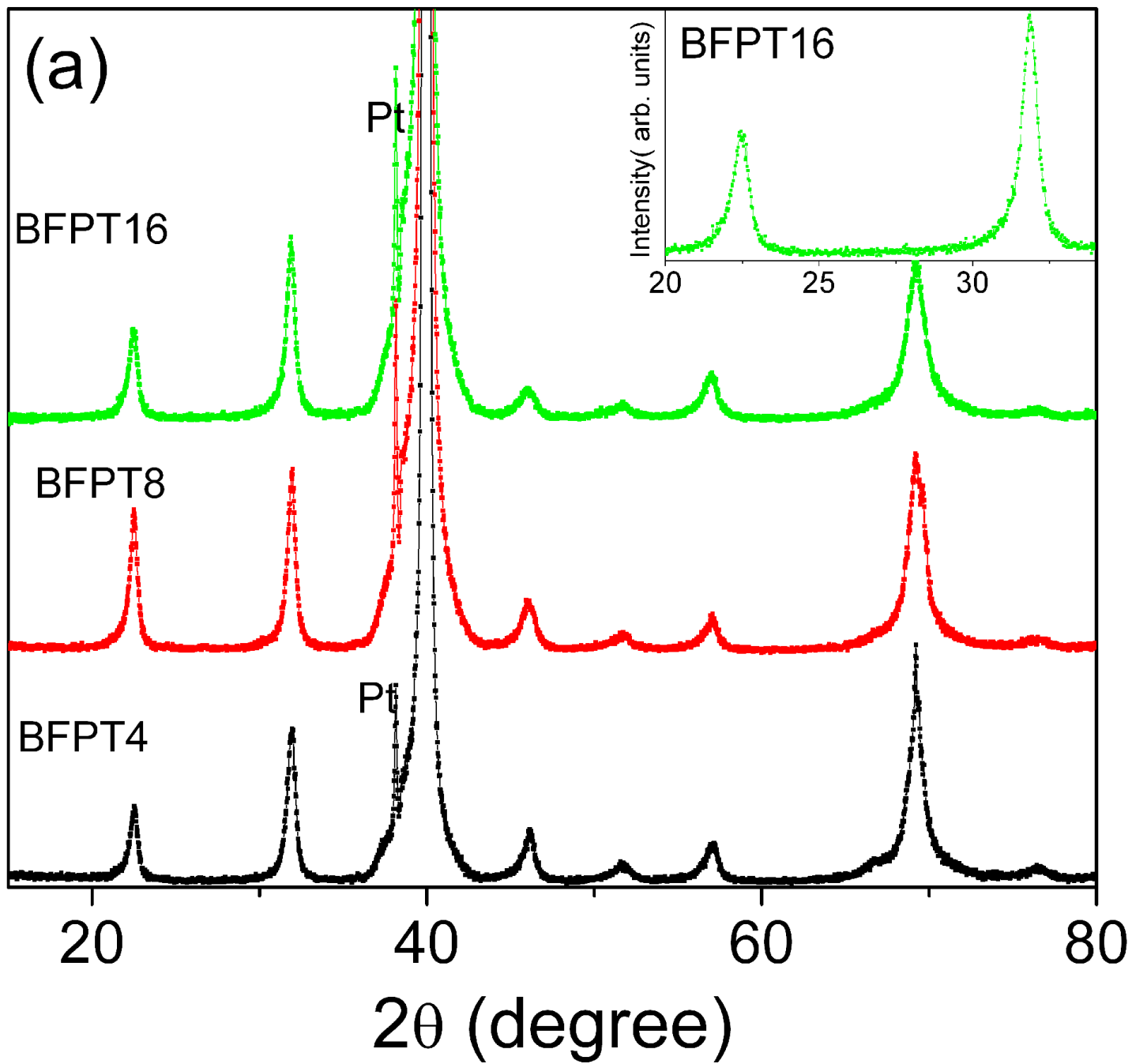

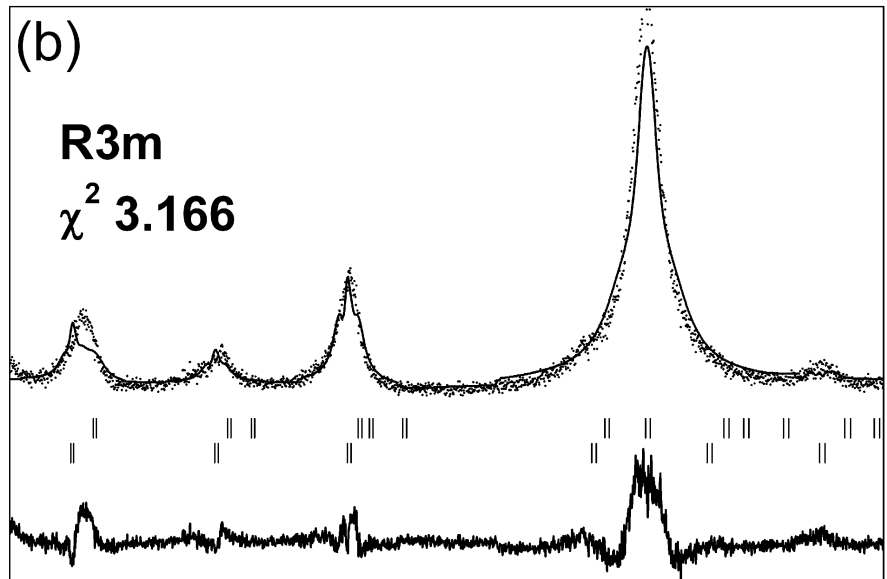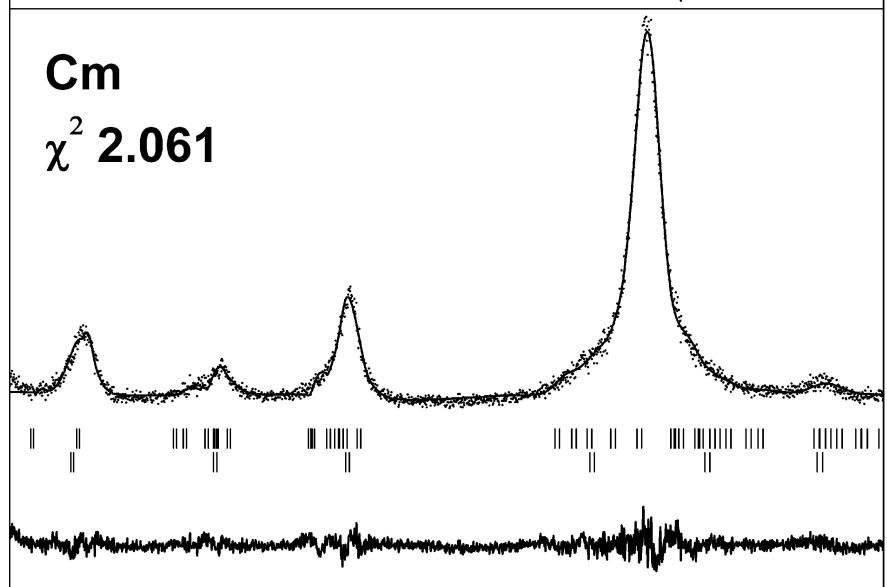

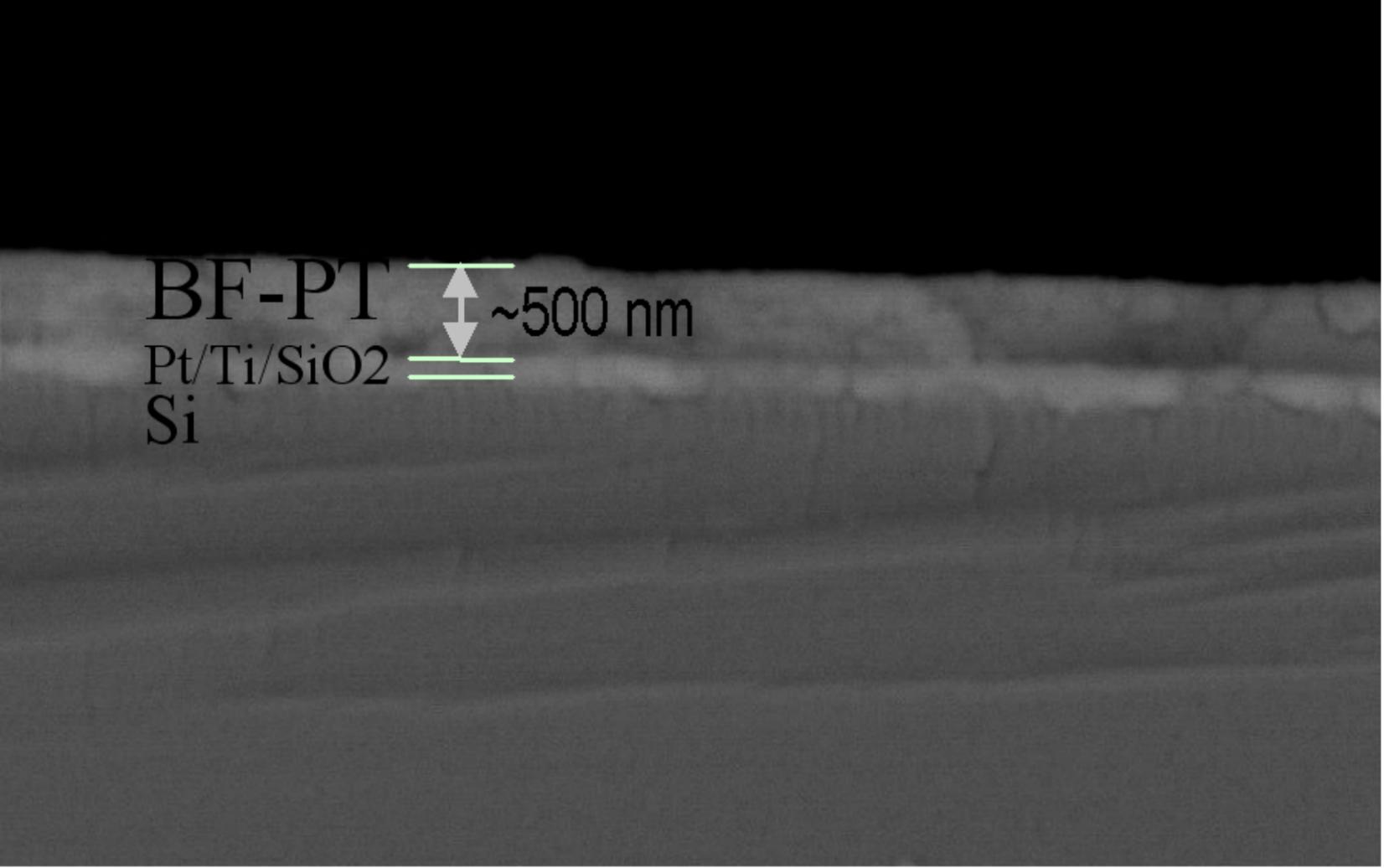

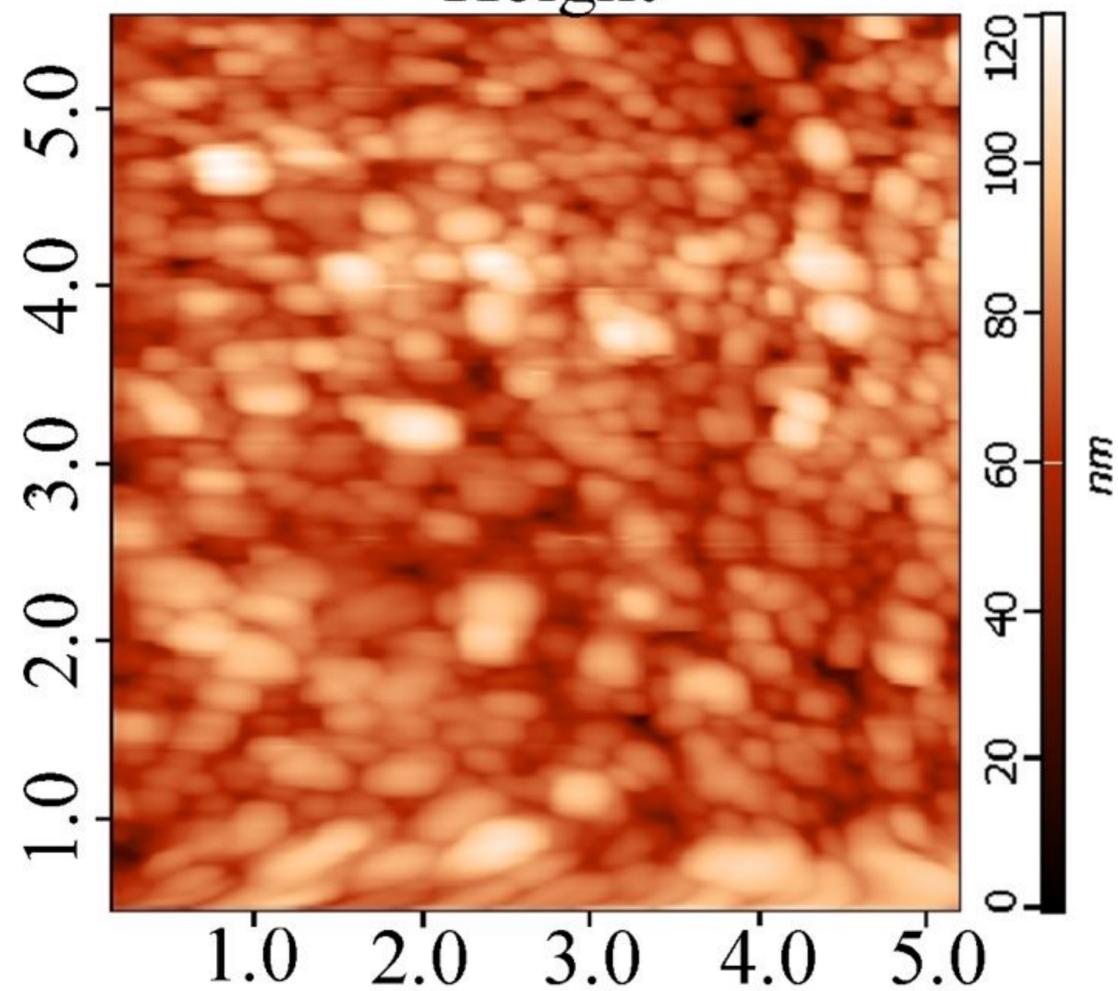
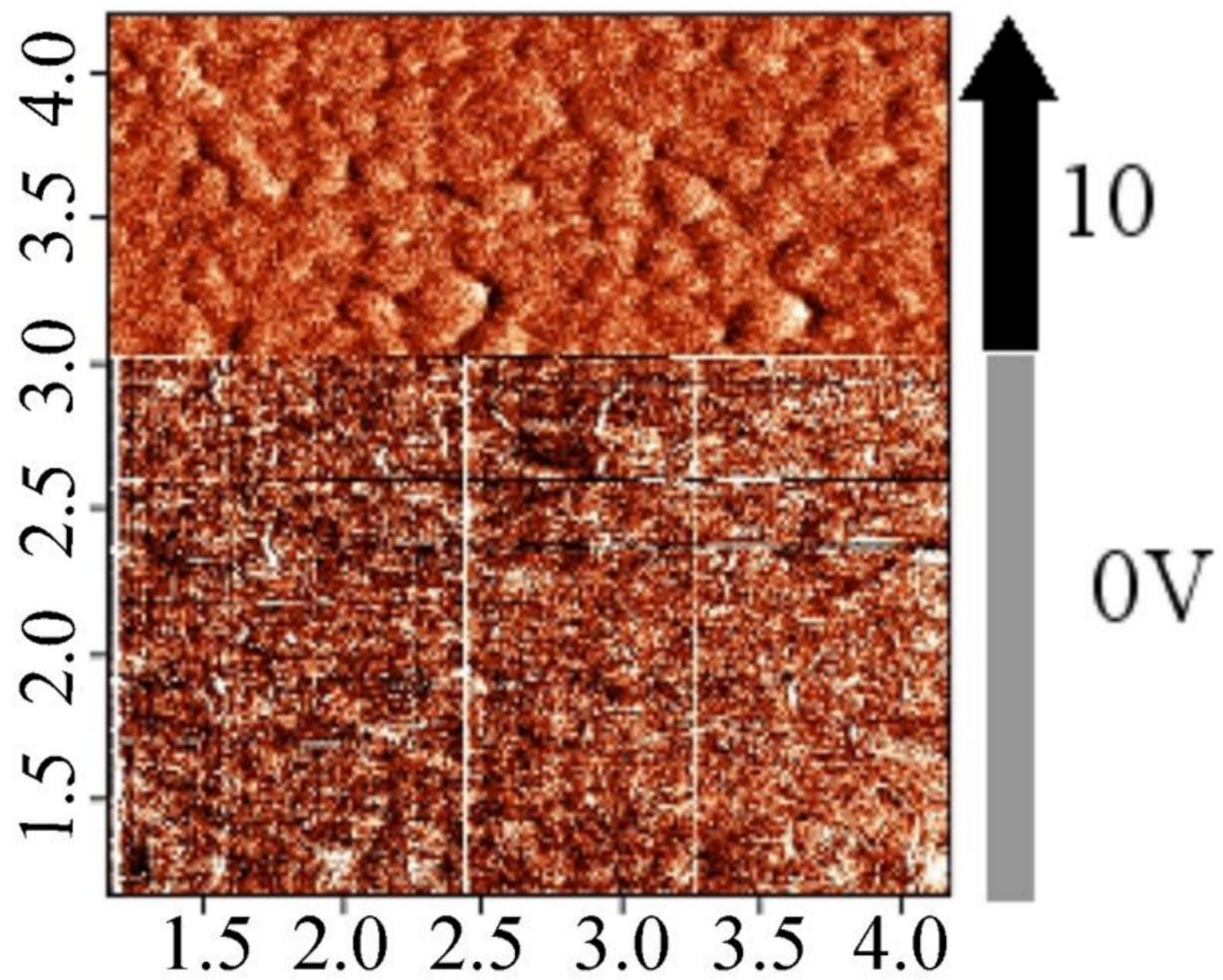
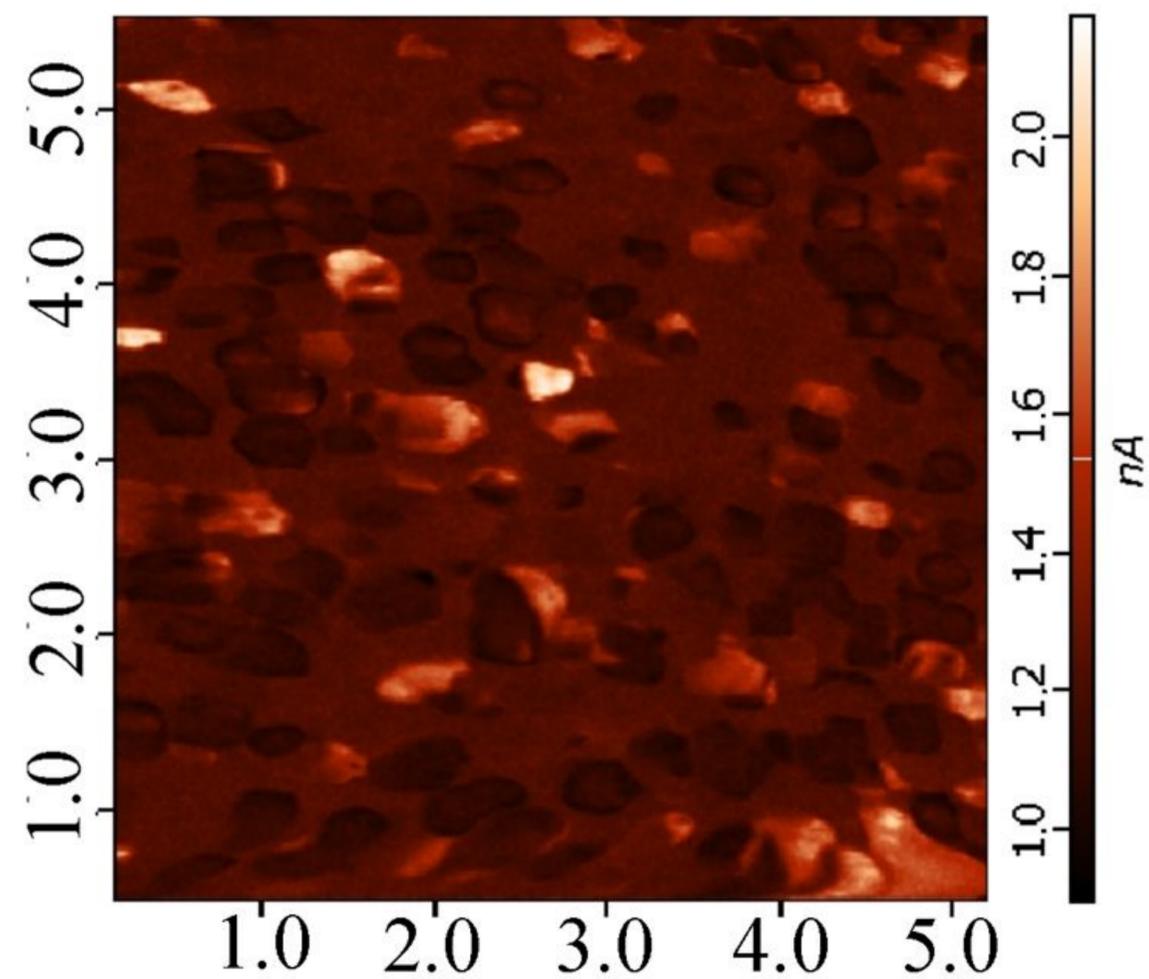
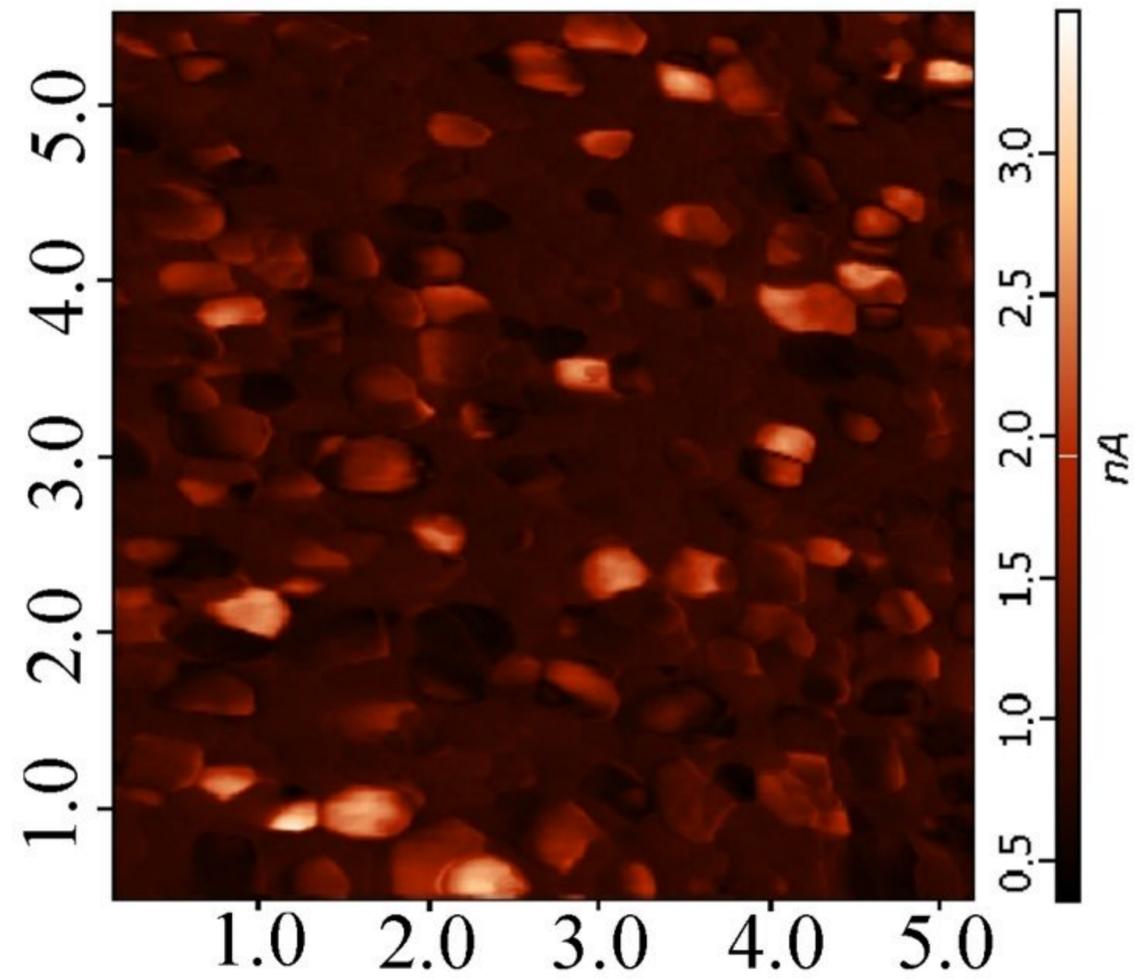

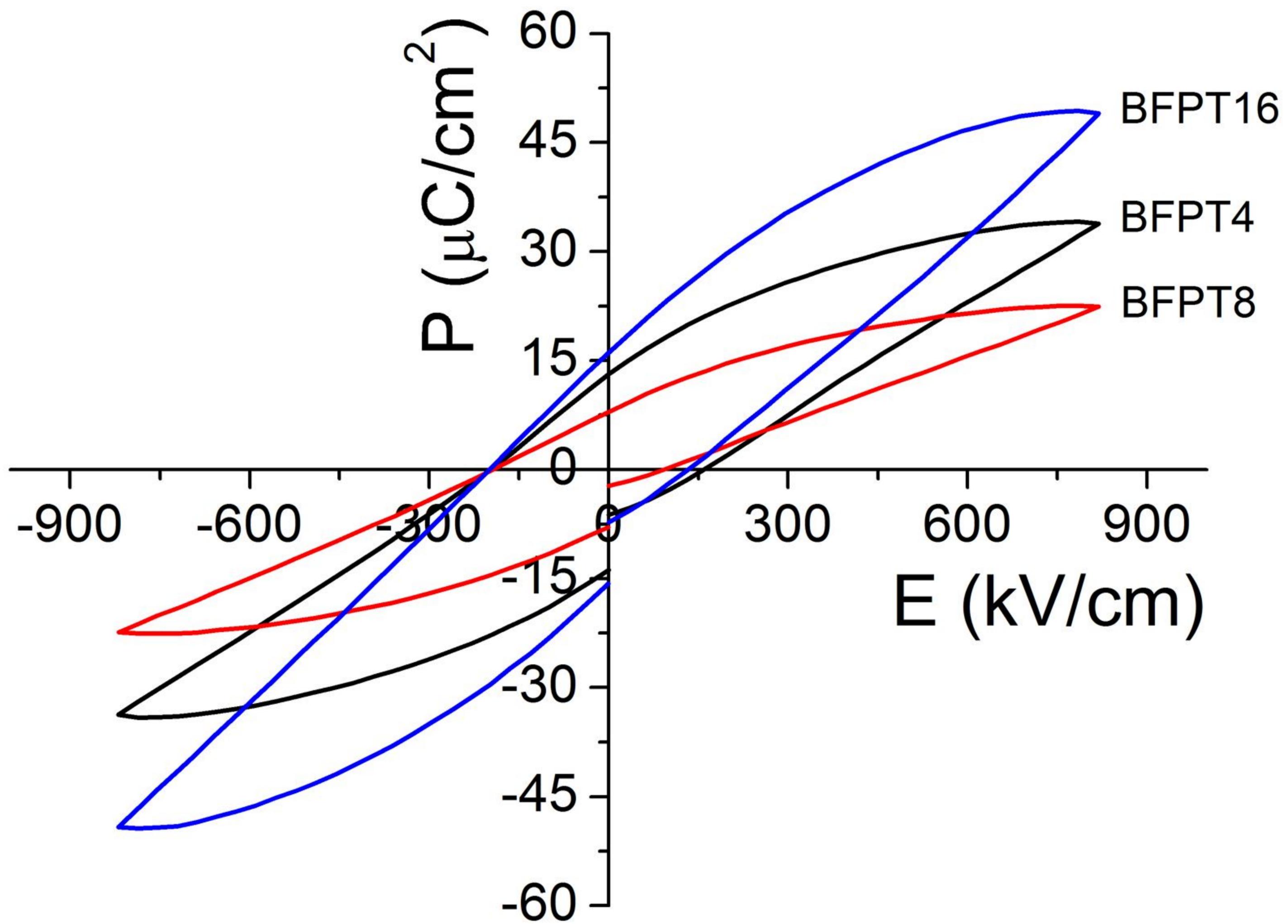

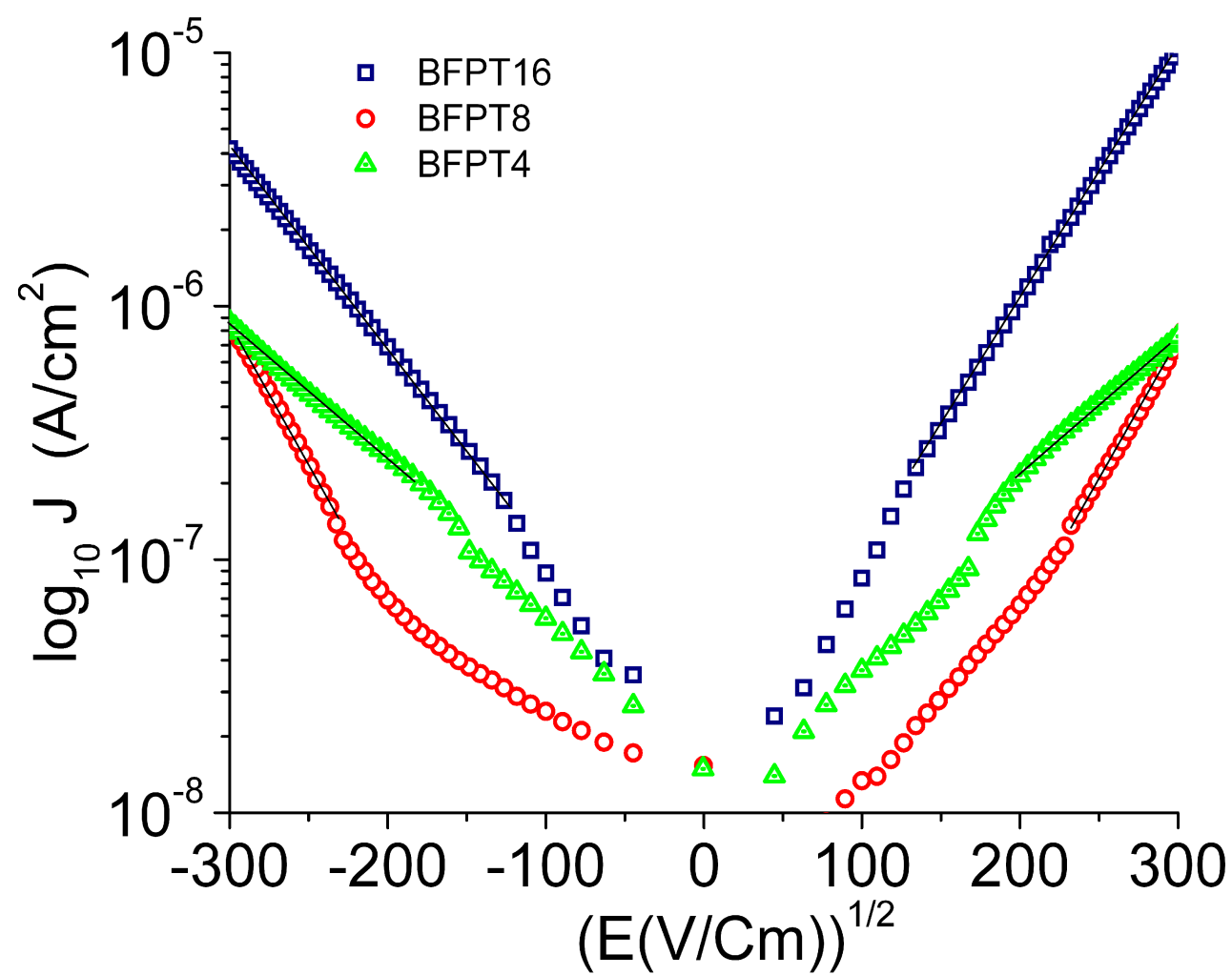

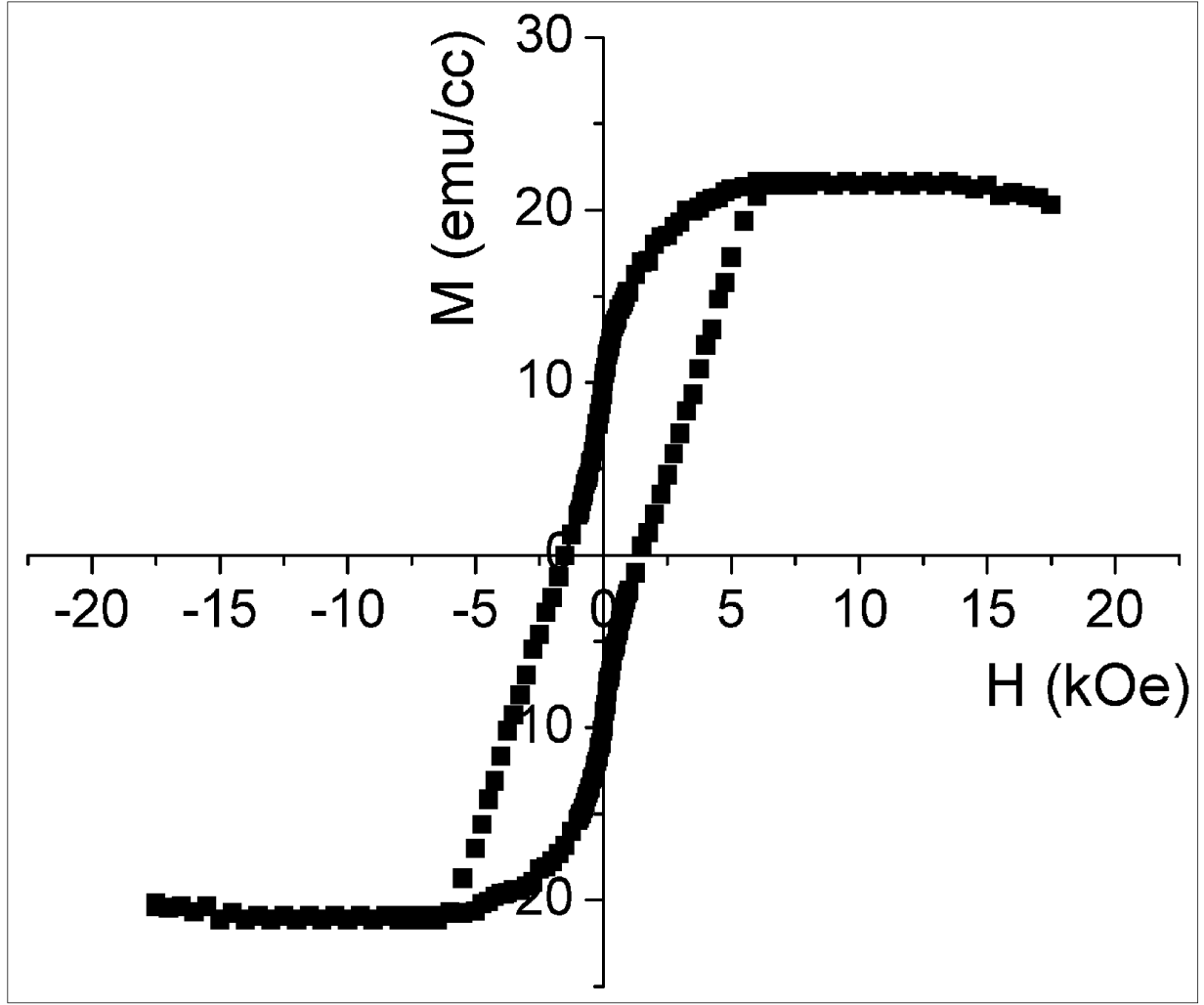